\documentclass[10pt]{article}

\usepackage{amsfonts,amssymb,amsmath,latexsym}
\usepackage{amsthm}
\usepackage{fullpage}

\newtheorem{theorem}{Theorem}[section]

\newcommand{\braket}[2]{\ensuremath{\langle #1 | #2 \rangle}}
\newcommand{\Qcont}{\ensuremath{\tilde{Q}}}
\newcommand{\px}{\ensuremath{\psi_{x}^{t}}}
\newcommand{\py}{\ensuremath{\psi_{y}^{t}}}
\newcommand{\ddt}{\ensuremath{\frac{d}{dt}}}
\newcommand{\ket}[1]{\left\vert{#1}\right\rangle}
\newcommand{\braUket}[3]{\ensuremath{\langle #1 | #2 | #3 \rangle}}
\newcommand{\braddtket}[2]{\braUket{#1}{\ddt}{#2}}
\newcommand{\ceil}[1]{\ensuremath{\left\lceil#1\right\rceil}}
 \newcommand{\twothirds}{\frac{2}{3}}
\newcommand{\Real}{\text{Re}}
\newcommand{\Imag}{\text{Im}}

\begin{document} 
\title{\LARGE \bf Adversary lower bounds in the Hamiltonian oracle model
\footnote{This manuscript was written in 2007.  The section on generalising to negative weights was added in 2011 at the suggestion of Troy Lee.}}

\author{David Yonge-Mallo}
\date{}

\maketitle \thispagestyle{empty}

\vspace{-2ex}
\begin{abstract}
In this note, we show that quantum lower bounds obtained using the adversary
method hold in the Hamiltonian oracle model.
\end{abstract}

\section{Introduction}

The adversary method is one of the two main techniques for proving lower
bounds in the quantum query model (the other being the {\em polynomial
method}).  It is an extremely versatile method with several equivalent
formulations which has been used to obtain good lower bounds for a 
variety of functions.  It can be understood in terms of weight schemes
\cite{ambainis2006:polynomial_vs_quantum_query, Zhang2005}, via
semidefinite programming and spectral analysis \cite{Barnuma}, or
through Kolmogorov complexity \cite{Laplante2003}.  All of these
formulations have been shown to be equal both in their power and in
their limitations \cite{Spalek2006}.  Later, an extension of the 
adversary method was introduced which allows the use of negative
weights and removes some of the limitations of the method \cite{Hoyera}.

\section{Discrete oracles, fractional oracles, and Hamiltonian oracles}

Suppose that we wish to compute some function 
$f : \{0,1\}^{N} \mapsto \{0,1\}$,
given the input variables $x = x_{1}x_{2}\cdots x_{N}$, using a quantum
algorithm.  The state of the algorithm
at any time $t$, on the input string $x$, may be written
in terms of a set of basis states $\ket{j,k}$ such that the 
first $\ceil{\log N}$ qubits $j$ range over the indices of the variables:
\[ \ket{\px} = \sum_{j,k}\alpha_{j,k}\ket{j,k} \]

In the conventional {\em (discrete) quantum query model}, access to the variables is 
allowed only through a discrete oracle, which can be queried with index $j$ to 
obtain the value of the variable $x_{j}$. 
The query complexity of any particular algorithm computing $f$ is the number of
queries made by that algorithm, and the query complexity of the function $f$ 
itself is the minimum query complexity of any algorithm computing $f$.
In this model, we 
typically\footnote{We could also have defined the query so that it maps a
basis state $\ket{j,b,k}$ to $\ket{j,b \oplus x_{j},k}$.  The two formulations
are essentially equivalent.}
define the query transformation $Q_{x}$ so that 
the basis state $\ket{j,k}$ queries the variable $x_{j}$, and gains a negative
phase if $x_{j} = 1$.  
Then the query maps $\ket{j,k}$ to $(-1)^{x_{j}}\!\ket{j,k}$, that is,
\[
Q_{x}\!\ket{j,k} =
\begin{cases}
~\ket{j,k} & \text{if $x_{j} = 0$,}\\
-\ket{j,k} & \text{if $x_{j} = 1$.}
\end{cases}
\]

In addition to queries, a discrete quantum query algorithm can also perform arbitrary 
unitary transformations that do not depend on the input string $x$.  An algorithm that
makes $T$ discrete queries ($T$ is an integer) is just a sequence of operations 
alternating between arbitrary unitary transformations and queries:
\[ U_{0}, Q_{x}, U_{1}, Q_{x}, U_{2}, Q_{x}, \ldots, Q_{x}, U_{T-1}, Q_{x}, U_{T} \]

The sequence is applied to the initial state $\ket{\psi^{0}}$ (which is independent
of the input $x$) to produce the final state $\ket{\psi_{x}^{T}}$, which is measured
by the algorithm to produce the output.  If the output is correct with probability
at least $\twothirds$, we say that the algorithm computes $f$ with bounded error.

The {\em fractional quantum query model} generalizes the discrete model by allowing
fractions of an oracle query to be made.  For integer $M$, the fractional 
query $Q_{x}^{1/M}$ maps $\ket{j,k}$ to $\left(e^{-i\pi/M}\right)^{x_{j}}\ket{j,k}$.
An algorithm in this model is a sequence of operations alternating between 
arbitrary unitary transformations and such fractional queries:
\[ U_{0/M}, Q^{1/M}_{x}, U_{1/M}, Q^{1/M}_{x}, U_{2/M}, Q^{1/M}_{x}, \ldots, Q^{1/M}_{x}, U_{T-1/M}, Q^{1/M}_{x}, U_{T} \]

The {\em Hamiltonian oracle model}, introduced in \cite{farhi1998:analog},
results from taking the limit $M \rightarrow \infty$ in the fractional query
model.  It is thus a continuous-time generalization of the discrete query
model (see \cite{mochon2006:hamiltonian_oracles, farhi2007:NAND_trees}).  
In this model, the state of a quantum algorithm $\ket{\px}$ evolves 
according to the Schr{\"o}dinger equation
\[
i\ddt\!\ket{\px} = H_{x}(t)\!\ket{\px}
\]
where $H_{x}(t)$ is the Hamiltonian of the algorithm.  The algorithm
starts in the initial state $\ket{\psi^{0}}$ and evolves for a time
$T$ to reach the final state $\ket{\psi_{x}^{T}}$.  The query 
complexity of a function $f$ is then the mininum time $T$ needed to 
compute $f$.

The Hamiltonian $H_{x}(t)$ may be decomposed into two parts, a
Hamiltonian oracle $H_{Q}(x)$ that depends on the input string $x$ but is 
independent of time, and a driver Hamiltonian $H_{D}(t)$ that depends
on the time $t$ but is independent of the input.  
(Thus, the Hamiltonian oracle corresponds to the oracles calls and
the driver Hamiltonian corresponds to the arbitrary unitary transformations
in the discrete query model.)
To be as general as possible, we can write the combined Hamiltonian, 
on the input string $x$, as
\[
H_{x}(t) = g(t)H_{Q}(x) + H_{D}(t)
\]
for some $|g(t)| \leq 1$.  

The Hamiltonian oracle $H_{Q}(x)$ has the form
\[
H_{Q}(x) = \sum_{j = 1}^{N} H_{j}(x)
\]
where each $H_{j}$ operates on an orthogonal subspace $V_{j}$.  That is,
writing $P_{j}$ as the projection onto $V_{j}$, we have
$H_{j} = P_{j}H_{j}P_{j}$.  We also assume that $||H_{j}|| \leq 1$.  
For each $j$, there are two possible operators $H_{j}^{(x_{j})}$,
corresponding to $x_{j} = 0$ and $x_{j} = 1$.  

To simulate the fractional or discrete query model using the 
Hamiltonian query model, let $H_{j}$ be the matrix with 
$\pi \cdot x_{j}$ in the $j$-th row and column, and zeroes elsewhere.  
Then each $H_{j}$ operates on an orthogonal subspace.  
Note that $H_{Q}(x)$ is simply the matrix with
the string $x$ on the first $N$ entries of the diagonal, multiplied 
by $\pi$, and zeroes everywhere else.  
If we now choose $g(t) = 1$ and $H_{D}(t) = 0$ and 
evolve the basis state $\ket{j,k}$ for a time $1/M$, 
the result will be the state $\left(e^{-i\pi/M}\right)^{x_{j}}\ket{j,k}$,
which simulates an oracle call.  
Likewise, an arbitrary unitary $U$ that is independent of the input may 
be simulated by setting $g(t) = 0$ and choosing $H_{D}(t)$ appropriately.

\section{The adversary method}\label{sec:method}

The primary idea behind the adversary method is that if an algorithm
computes a function, then it must be able to distinguish between inputs
that map to different outputs.  A certain amount of information about 
the inputs is required to distinguish them, and thus one may obtain
lower bounds for the number of queries required to compute a function
by upper bounding the amount of information revealed in each query.

There are several equivalent formulations of the adversary method.
We describe the spectral formulation below because it is convenient.
The proof below is essentially a continuous version of the proof
from \cite{Hoyer2005}.

Consider a pair of inputs $x$ and $y$ such that $f(x) = 0$ and $f(y) = 1$.
As above, we write $\ket{\px}$ to denote the state of the
quantum algorithm on input $x$ at time $t$, and similarly for $y$.
If the algorithm finishes after $T$ queries, we would like 
$\ket{\psi_{x}^{T}}$ and $\ket{\psi_{y}^{T}}$ to be easily
distinguishable, or equivalently, to have a small inner product.
To distinguish the two states correctly with error probability at
most $\epsilon$, we require 
$|\braket{\psi_{x}^{T}}{\psi_{y}^{T}}| \leq \epsilon'$, 
where $\epsilon' = 2\sqrt{\epsilon(1-\epsilon)}$.
(This is known as the Ambainis output condition \cite{Barnuma}.)
We can use this idea to define a {\em progress measure} using 
the inner products between all pairs of inputs.  

To capture the fact that some pairs of inputs are more difficult to 
distinguish than others, we assign a {\em weight} to each pair.  
To do so, we define a {\em spectral adversary matrix} $\Gamma$, 
which is a symmetric $2^{N} \times 2^{N}$ matrix of non-negative real 
values such that $\Gamma[x,y] = 0$ whenever $f(x) = f(y)$. 
(The following argument actually holds for the general adversary method, 
and not just for the non-negative method; see Section~\ref{sec:general} below.)
Let $\delta$ be a fixed principal eigenvector of $\Gamma$.
We now define the progress measure to be 
\[ w^{t} = \sum_{x,y} \Gamma[x,y] \cdot \delta[x] \cdot \delta[y] \cdot 
\braket{\px}{\py} \]
where $\Gamma[x,y]$ is the entry corresponding to the $x^{\mathrm{th}}$ row 
and $y^{\mathrm{th}}$ column of $\Gamma$, and similarly $\delta[x]$ is the 
$x^{\mathrm{th}}$ entry of $\delta$.  We also define, for $1 \leq i \leq N$, 
a related family of matrices
\[
\Gamma_{i}[x,y] = 
\left\{
\begin{array}{ll}
\Gamma[x,y] & x_{i} \neq y_{i},\\
0 & x_{i} = y_{i}.
\end{array}
\right.
\]

Let $\Qcont_{2}(f)$ denote the bounded-error query complexity in the 
Hamiltonian oracle model for $f$, and write $\lambda(M)$ for the
spectral norm of a matrix $M$.  
The spectral version of the adversary theorem in the Hamiltonian
oracle model is essentially the same as in the discrete query
model.
\begin{theorem}
For any adversary matrix $\Gamma$ for $f$,
\[
\Qcont_{2}(f) = \Omega\left( \frac{\lambda(\Gamma)}{\max_{j}\lambda(\Gamma_{j})} \right).
\]
\end{theorem}

The algorithm starts in an initial state $\ket{\psi^{0}}$ which is 
independent of the input, and thus the initial value of the progress 
measure is
\[ w^{0} = \sum_{x,y} \Gamma[x,y] \cdot \delta[x] \cdot \delta[y] =
\delta^{T} \Gamma \delta = \lambda(\Gamma). \]

To lower bound the time required for the algorithm to succeed, we 
upper bound the change in the progress measure.  We can do this by
taking its derivative with respect to time.  First, note that
\begin{eqnarray}
\ddt \braket{\px}{\py}\braket{\py}{\px} & = &
\braket{\px}{\py}\left(\ddt \braket{\py}{\px}\right) + 
\left(\ddt \braket{\px}{\py}\right)\braket{\py}{\px} \nonumber \\
& = & 
2 \Real\!\left[ \braket{\px}{\py} \left(  
\braddtket{\py}{\px} + \left(\braddtket{\px}{\py}\right)^{*} \right) \right] \nonumber\\
& = & 
2 \Real\!\left[ -i\braket{\px}{\py} 
\left( \braUket{\py}{H_{x}(t)}{\px} - \braUket{\py}{H_{y}(t)}{\px} \right) 
\right] \nonumber\\
& = & 
2 \Imag\!\left[ \braket{\px}{\py} 
\braUket{\py}{(H_{x}(t) - H_{y}(t))}{\px} \right] \nonumber 
\end{eqnarray}

Next, we can upper bound the change in the magnitude of the 
inner products between the algorithm states corresponding to each 
pair of inputs $x$ and $y$:
\begin{eqnarray}
\ddt |\braket{\px}{\py}|
& = & \ddt 
\sqrt{\braket{\px}{\py}\braket{\py}{\px}} \nonumber\\
& = & \frac{1}{2 |\braket{\px}{\py}|} 
\ddt \braket{\px}{\py}\braket{\py}{\px} \nonumber\\
& = & \frac{1}{|\braket{\px}{\py}|} 
\Imag\!\left[ \braket{\px}{\py} 
\braUket{\py}{(H_{x}(t) - H_{y}(t))}{\px} \right] \nonumber\\
& \leq & 
\left|\braUket{\py}{(H_{x}(t) - H_{y}(t))}{\px}\right| \label{eqn:w_derivative}
\end{eqnarray}

We can rewrite the difference $H_{x}(t) - H_{y}(t)$ as:  
\begin{eqnarray*}
H_{x}(t) - H_{y}(t)
& = & \left\{g(t)H_{O}(x) + H_{D}(t)\right\} - 
\left\{g(t)H_{O}(y) + H_{D}(t)\right\} \\
& = & \sum_{j : x_{j} \neq y_{j}} g(t)\left(H_{j}^{(x_{j})} - H_{j}^{(y_{j})}\right) 
\end{eqnarray*}

This shows that that the progress measure $w^{t}$ does not depend on 
the driver Hamiltonian $H_{D}(t)$.  
Now let $\Delta_{j} = g(t)\left(H_{j}^{(x_{j})} - H_{j}^{(y_{j})}\right)$,
and note that $||\Delta_{j}|| \leq 2$ for all $j$.
Substituting into Equation (\ref{eqn:w_derivative}), we have:
\begin{eqnarray}
\ddt |\braket{\px}{\py}|
& \leq & \left| \sum_{j : x_{j} \neq y_{j}} 
\braUket{\py}{P_{j}\Delta_{j}P_{j}}{\px}
\right| \nonumber\\
& \leq & 
\sum_{j : x_{j} \neq y_{j}} 
\left| 
\braUket{\py}{P_{j}\Delta_{j}P_{j}}{\px}
\right| \nonumber\\
& \leq & 2 
\sum_{j : x_{j} \neq y_{j}}
\left|\left| P_{j} \ket{\px} \right|\right| \cdot 
\left|\left| P_{j} \ket{\py} \right|\right| \nonumber
\end{eqnarray}

Let $\beta_{x,j} = \left|\left| P_{j} \ket{\px} \right|\right|$ denote the
absolute value of the amplitude querying $x_{j}$ at time $t$, and note
that $\sum_{j} \beta_{x,j}^{2} = 1$.  
We define an auxiliary vector $a_{j}[x] = \delta[x]\beta_{x,j}$
which has the property that 
$\sum_{j}|a_{j}|^2 = 
\sum_{j}\sum_{x}\delta[x]^{2}\beta_{x,j}^{2} = 
\sum_{x}\delta[x]^{2} \sum_{j}\beta_{x,j}^{2} = 
\sum_{x}\delta[x]^{2} = 1$.

Finally, we can upper bound the derivative of the magnitude of the
progress measure as follows:
\begin{eqnarray*}
\ddt |w^{t}| & = & 
\sum_{x,y} \Gamma[x,y] \cdot \delta[x] \cdot \delta[y] \cdot \left|\ddt\braket{\px}{\py}\right|\\
& \leq & 2 \sum_{x,y} \sum_{j} 
\Gamma[x,y] \cdot \delta[x] \cdot \delta[y] \cdot \beta_{x,j} \cdot \beta_{y,j}\\
& = & 2 \sum_{j} a_{j}^{T} \Gamma_{j} a_{j}\\
& \leq & 2 \sum_{j} \lambda(\Gamma_{j})|a_{j}|^{2}\\
& \leq & 2 \max_{j} \lambda(\Gamma_{j}) \cdot \sum_{j}|a_{j}|^{2}\\
& = & 2 \max_{j} \lambda(\Gamma_{j})
\end{eqnarray*}

In order for the algorithm to succeed, we must have $w^{T} \leq \epsilon' w^{0}$.
Since we have $w^{0} = \lambda(\Gamma)$ and $\ddt |w^{t}| = 2 \max_{j} \lambda(\Gamma_{j})$,
we can integrate to obtain the theorem.

\section{Comparison with the FGG proof of the lower bound for parity}

When Farhi and Goldstone introduced the Hamiltonian oracle model 
in \cite{farhi1998:analog} and used it to prove a lower bound on a 
continuous-time version of Grover's search, they referred to their technique 
as the ``analog analogue'' of the BBBV method \cite{Bennett1997}.
As the discrete query adversary method is an extension of the BBBV
method, the Hamiltonian oracle version of the adversary may be seen
as an extension of the proof method introduced in \cite{farhi1998:analog},
and indeed, it is implicit in their and Gutmann's proof of the lower
bound for the parity problem in the Hamiltonian oracle model
\cite{farhi2007:NAND_trees}).  

In that paper, the progress measure used is
\begin{eqnarray*}
\left|\left| \ket{\px} - \ket{\py} \right|\right|^2 
& = &
\left(\ket{\px} - \ket{\py}\right)^{*}\left(\ket{\px} - \ket{\py}\right)\\
& = & 1 - \braket{\px}{\py} - \braket{\py}{\px} + 1\\
& = & 2 - 2 \Real\left[ \braket{\px}{\py} \right]
\end{eqnarray*}
and the derivative with respect to time of this progress measure
is essentially the same (up to a multiplicative factor of $\pm 2$) 
of the one used in this paper.  

\section{Addendum: Generalising to negative weights}\label{sec:general}

The argument in Section~\ref{sec:method} actually holds for the general
adversary method, and not just for the non-negative method.  The non-negative
method relies on the fact that an algorithm that computes a function must
distinguish between inputs that map to different outputs.  The general method
makes explicit use of the stronger condition that any such algorithm must
actually compute the function, by removing the restriction on the spectral
adversary matrix $\Gamma$ that its entries be real and non-negative.  Even with
this modification, the rate of change of the potential function, $\ddt
|w^{t}|$, is still upper bounded, as above.  That quantum lower bounds obtained
using the general adversary method hold in the Hamiltonian oracle model follows
from a version of the proof of Theorem 2 in \cite{Hoyera}.

\section*{Acknowledgment}

This research was done under the guidance of Richard Cleve at the Institute for
Quantum Computing.  The author would also like to thank Troy Lee for pointing
out that the argument generalises to negative weights as described in
Section~\ref{sec:general}.

\bibliographystyle{alpha}
\bibliography{adversary}
\end{document}